\documentclass[aps,pre,twocolumn,amsmath,amssymb]{revtex4-1}
\usepackage{color}
\usepackage{graphicx}
\usepackage{amsmath}
\usepackage[normalem]{ulem}
\begin{document}

\title{Friction weakening by mechanical vibrations: \\ a velocity-controlled process}

%% Notice placement of commas and superscripts and use of &
%% in the author list

%\author{Aauthor$^{1,2}$, Bauthor$^2$ \& LastAuthor$^2$}
\author{V. Vidal$^{1*}$, C. Oliver$^{2}$, H. Lastakowski$^{1}$, G. Varas$^{2}$ \& J.-C G{\'e}minard$^{1}$}\
\affiliation{$^{1}$ Universit\'e de Lyon, Laboratoire de Physique, ENS de Lyon, CNRS, F-69342 Lyon, France}
\affiliation{$^{2}$ Instituto de Fisica, Pontificia Universidad Cat{\'o}lica de Valparaiso, Av. Universidad 330, Valparaiso, Chile}

\date{\today}

%%%%%%%%%%%%%%%%%%%%%%%%%%%%%%%%%%%%%%%%%
\begin{abstract}
%For Nature, the abstract is really an introductory paragraph set
%in bold type.  This paragraph must be ``fully referenced'' and
%less than 180 words for Letters.  This is the thing that is
%supposed to be aimed at people from other disciplines and is
%arguably the most important part to getting your paper past the
%editors.  End this paragraph with a sentence like ``Here we
%show...'' or something similar.
Frictional weakening by vibrations was first invoked in the 70's to explain unusual fault slips and earthquakes, low viscosity during the collapse of impact craters or the extraordinary mobility of sturzstroms, peculiar rock avalanches which travels large horizontal distances. This mechanism was further invoked to explain the remote triggering of earthquakes or abnormally large landslides or pyroclastic flows runout. Recent experimental and theoretical work pointed out the velocity of vibration as the key parameter which governs frictional weakening in sheared granular media. Here we show that the grains mobility is not mandatory, and that the vibration velocity governs both granular and solid frictional weakening. The velocity threshold controlling the transition from stick-slip motion to continuous sliding is of the same order of magnitude, namely a hundred microns per second. It is linked to the roughness distribution of the asperities at the contact surface.
\end{abstract}
%%%%%%%%%%%%%%%%%%%%%%%%%%%%%%%%%%%%%%%%%

\maketitle

%Then the body of the main text appears after the intro paragraph.
%Figure environments can be left in place in the document.
%\verb|\includegraphics| commands are ignored since Nature wants
%the figures sent as separate files and the captions are
%automatically moved to the end of the document (they are printed
%out with the \verb|\end{document}| command. However, tables must
%be manually moved to the end of the document, after the addendum.

%\begin{figure}
%\caption{Each figure legend should begin with a brief title for
%the whole figure and continue with a short description of each
%panel and the symbols used. For contributions with methods
%sections, legends should not contain any details of methods, or
%exceed 100 words (fewer than 500 words in total for the whole
%paper). In contributions without methods sections, legends should
%be fewer than 300 words (800 words or fewer in total for the whole
%paper).}
%\end{figure}

{\it ``It is easier to further the motion of a moving body than to move a body at rest.''} This sentence written by Themistius (about A.D. 320-390) is the first record of friction in history~\cite{Sambursky62}. Since then, the frictional motion of a single body over a fixed substrate or of a sheared granular assembly revealed a wide variety of behaviors. At low shear velocity, the system experiences a stick-slip motion, with the alternance of loading phases (system at rest) and quick slip phases which release the energy. When increasing the shear velocity, a transition to continuous sliding motion is reported~\cite{Baumberger94,Marone98,Nasuno98,Baumberger06}. During catastrophic events such as earthquakes, landslides or pyroclastic flow, puzzling phenomena of frictional weakening were reported~\cite{Collins03,Lucas14,Xia13,Levy15}: friction decreases with the shear velocity. Melosh~\cite{Melosh79} first proposed in 1979 that vibrations due to particle collisions could temporarily reduce the normal stress, and thus decrease the shear stress threshold to trigger sliding motion~\cite{Melosh96}. This mechanism, initially called {\it vibrational fluidization} and later on {\it acoustic fluidization}~\cite{Melosh79,Melosh96}, was further sought to be at the origin of dramatic events triggered by external waves, such as earthquake remote triggering~\cite{Johnson05,Xia13}.
%invoked to explain the remote triggering of earthquakes~\cite{Johnson05,Xia13}.

%%%%%%%%%%%%%%%%%%%%%%%%%%%%%%%%%%%%%%%%%
% FIGURE 1
%%%%%%%%%%%%%%%%%%%%%%%%%%%%%%%%%%%%%%%%%
\begin{figure}[!b]
\begin{center}
\includegraphics[width=0.85\columnwidth]{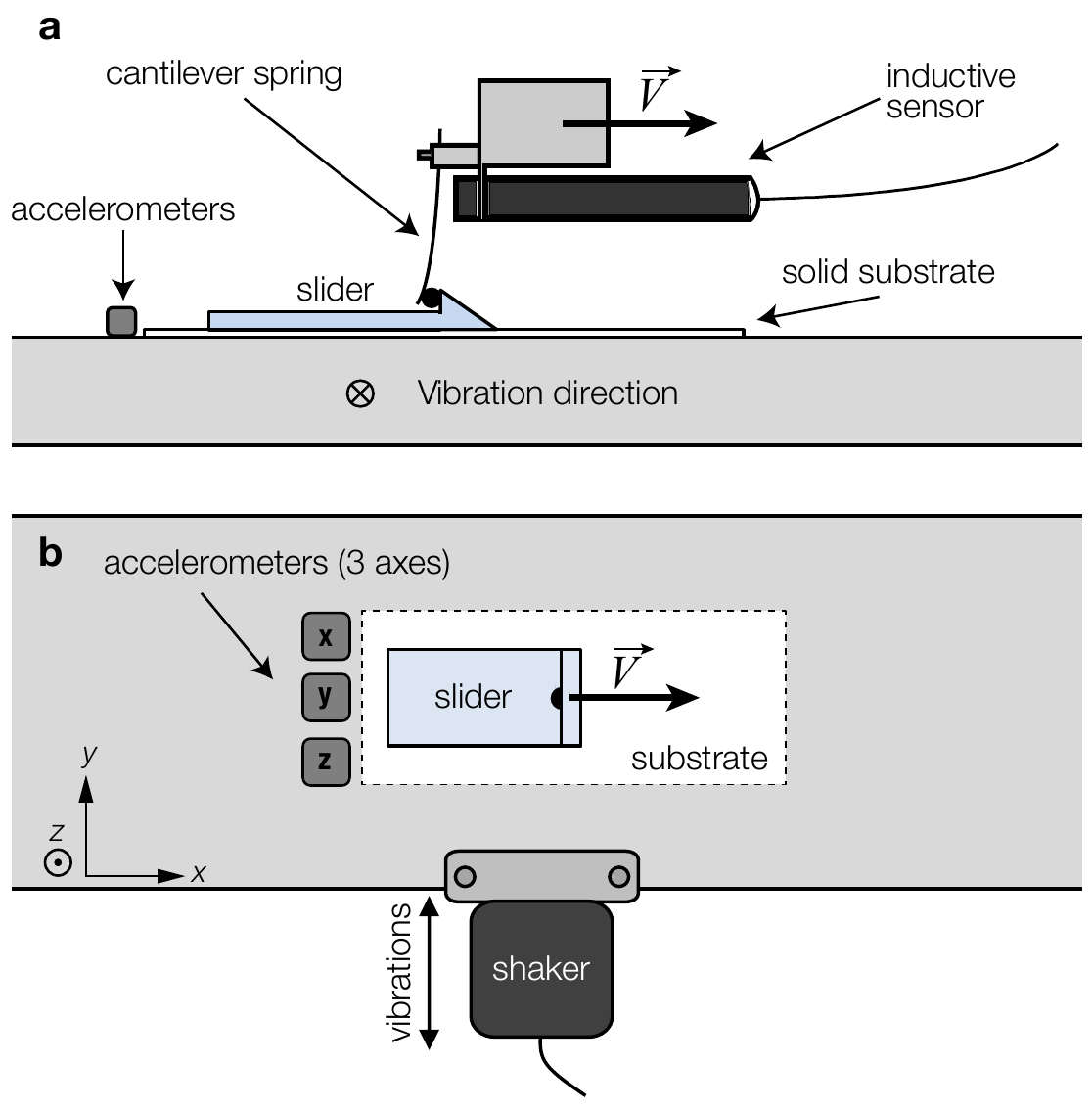}
\end{center}
\caption{\label{fig:expsetup} 
Sketch of the experimental setup. 
(a) Side view. 
%The slider (mass $m$) is pulled at constant velocity $V$ by a translational axis via a cantilever spring (stiffness $k$). An inductive sensor positioned behind the blade measures the distance from the cantilever and thus the force exerted over the blade. 
(b) Top view.  
%A shaker imposes horizontal vibrations in the transverse direction (along the $y$--axis). Three accelerometers are used to measure in-situ the vibrations amplitude and frequency. 
Different substrates are used in the experiment. In the case of granular material, the accelerometers are located at the bottom of the granular layer, aligned with the shaker.
}
\end{figure}

How does endogenous noise or external mechanical disturbances drastically affect the frictional properties? Many works have attempted to tackle this issue for the last decades. They have shown that vibrations reduce or even suppress friction~\cite{Rozman98,Gao98,Johnson08,Capozza09,Capozza11,Melhus12,Giacco12,Giacco15,Lastakowski15}. Interpretations were proposed based on a non-monotonic rheology curve, leading to instabilities and self-fluidization ~\cite{Dijksman11,Wortel16}, softening effect due to non-linearity at the grains contact~\cite{Jia11}, contact opening~\cite{DeGiuli15,Ferdowsi15} or sliding~\cite{DeGiuli17}. In single-block solid friction models, the vibration acceleration has often been stated at the parameter governing the transition between stick-slip motion and continuous sliding, with a threshold equal to the gravitational acceleration~\cite{Giacco12}.
In a recent work, Lastakowski et al.~\cite{Lastakowski15} pointed out that the {\it vibration velocity}, and not the acceleration, is the parameter governing the frictional weakening in granular assemblies. This result is independent of the slider velocity, contrary to what is expected from the classical rate-and-state heuristic model ~\cite{Gu84,Rice86,Marone98,Baumberger06}. Surprisingly, the sheared granular exhibits a transition between stick-slip motion and continuous sliding for very low values of the vibration velocity, around 100~$\mu$m/s, independently of most parameters which can be varied in the system (slider velocity, grain size and material, granular layer thickness, ...). A recent microscopic model based on sliding contacts under vibrations in a granular assembly successfully explains the dependence of the transition on the vibration velocity~\cite{DeGiuli17}. However, important questions arise: Is the grains mobility mandatory to experience this transition? Is vibration velocity the parameter governing the transition in solid friction too?
Here we address these questions by studying experimentally solid (paper-paper) friction under harmonic vibrations. 

{\it Experimental setup --}
The setup is similar to the one used by Lastakowski et al.~\cite{Lastakowski15}  (Fig.~\ref{fig:expsetup}) to further compare the frictional weakening under vibrations in both solid and granular friction. It consists of a slider, made of plexiglas, of length 9~cm and width 6~cm, pulled over a fixed substrate by means of a cantilever spring (metallic blade of stiffness $k$) (Fig.~\ref{fig:expsetup}). A steel spherical ball is glued at the front of the slider to ensure a punctual contact so that no torque is applied to the slider. The blade is mounted on a translational stage (Schnaefler Technologies Sechnr) moving at constant velocity $V$. A DC motor (Crouzet, 5~N.m, 17~W) coupled with differents reduction gears (Crouzet 1.04, 10, 100 RPM) allows to achieve velocities between 18-7700 $\mu$m/s. An inductive sensor (Baumer, IPRM 12I9505/S14) measures the blade deflection at a rate of 2~kHz. From the variations of the blade deflection in time, we derive the instantaneous force $F$ applied to the slider, and denote $F^*=F/mg$ the dimensionless force, where $m \simeq 22 \pm 5$~g is the slider mass and $g=9.81$~m.s$^{-2}$ the gravitational acceleration. 

Paper-paper friction is investigated by using two sets of samples: smooth printer paper (Inapa tecno copy-/laser pro laser, 80~g/m$^3$, white) and rough drawing paper (Canson$^\circledR$  Papier à dessin blanc C à grain, 180~g/m$^2$, rough surface). 
%To ensure reproducible experiments, both the surface below the slider and upon the substrate are prepared according to the following protocol. First, the samples are cut carefully, without touching the surface to avoid contaminating and modifying its properties. A first sheet of large dimensions (length 21~cm, width 9~cm) is stuck on the solid substrate; a second sheet cut at the slider's dimensions is stuck below the slider, ensuring a paper-paper frictional contact (Fig.~\ref{fig:expsetup}). For each given set of parameters (translation velocity $V$, spring stiffness $k$ and vibrations amplitude and frequency), three runs are performed with the same paper samples to check the reproducibility. The samples are then removed from both the substrate and the slider bottom face and replaced by new ones, to avoid damage and sample aging due to repetitive friction of the surfaces. The first experiment with the new samples is made with the previous set of parameters, to test reproducibility from one sample to the other, then parameters are changed, three runs are performed, and so on. 
%
Reproducibility is systematically checked by performing different experiments for each given set of parameters.
To avoid possible variations due to atmospheric conditions, the whole experiment is set inside a large box  of controlled temperature $T$ and humidity $R_H$. For all experiments on paper-paper friction, $T=35 \pm2^\circ$C and $R_H=20 \pm 2$\%. Results for granular material are inferred from the analysis of the previous experimental data of Lastakowski et al.~\cite{Lastakowski15}.

Vibrations are imposed to the whole experiment by a shaker (Br{\"u}el \& Kj\ae r, type 4810 + amplifier 2706) clamped on the aluminum frame. It applied horizontal harmonic vibrations perpendicular to the slider's direction of motion, along the $y$--axis (Fig. \ref{fig:expsetup}). The vibrations amplitude $A$ and frequency $\omega$ are measured {\it in situ} close to the slider, at the surface of the solid substrate, by three accelerometers (Dytran Instrument, model $\#$ 3035BG) getting the three components of acceleration. Before performing any experiment, we checked that the local acceleration is correctly oriented in the $y$--direction, and constant over a region large enough to include the slider motion. 
%We avoid resonances due to the experimental frame. Note that for the experiments on granular friction, the local acceleration was measured by sticking the accelerometers at the bottom of the granular layer, below the slider trajectory.

%Additional SEM and AFM measurements are performed to investigate the link between the properties of frictional weakening and the roughness of the solid substrate. Scanning Electron Microscopy (SEM) images were adquired on a Supra 55, VP Zeiss. Atomic Force Microscopy (AFM) measurements were performed on a commercial apparatus (NanoWizard$^\circledR$ 4 AFM, JPK Instruments) on paper samples at different spatial resolution ($256 \times 256$~pixels for a spatial extent going from $0.352 \times 0.352$ to $100 \times 100$~$\mu$m). 

%%%%%%%%%%%%%%%%%%%%%%%%%%%%%%%%%%%%
% FIGURE 2
%%%%%%%%%%%%%%%%%%%%%%%%%%%%%%%%%%%%
\begin{figure}[!t]
\begin{center}
\includegraphics[width=0.9\columnwidth]{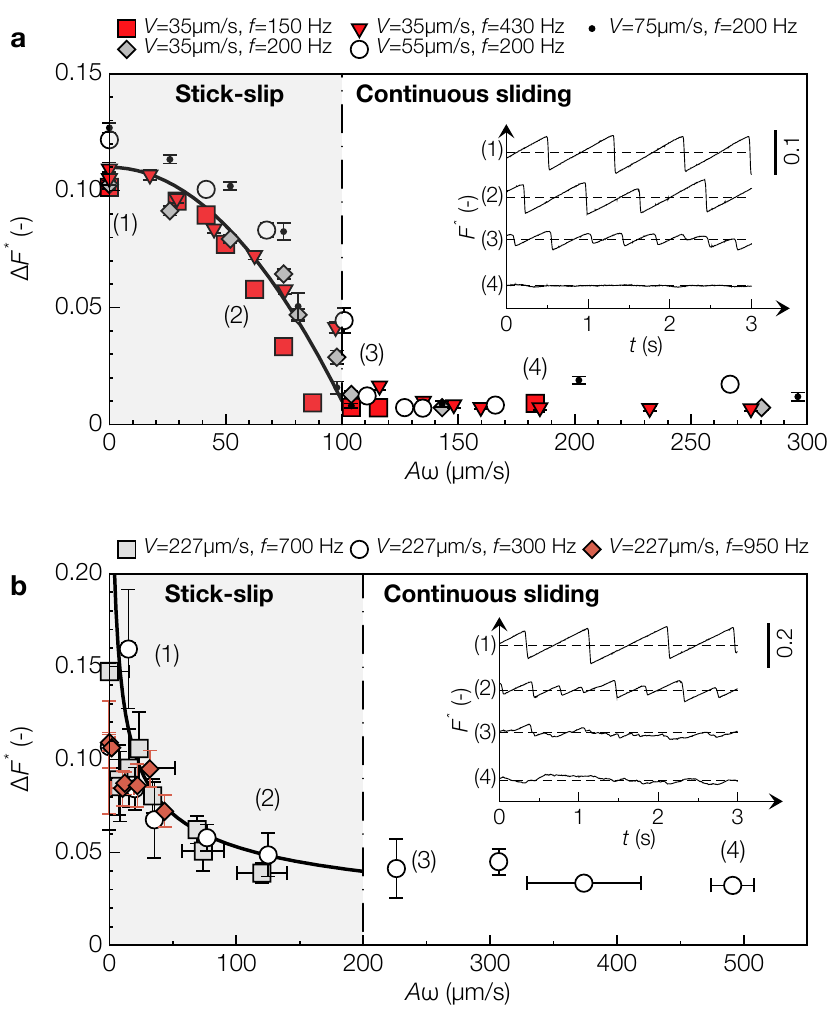}
\end{center}
\caption{\label{fig:DF}
Normalized force applied on the slider, $F^*$ for (a) granular and (b) solid (paper-paper, Inapa) friction [black lines are guides to the eye]. {\it Insets:} examples of normalized force as a function of time when increasing the vibration velocity, $(A\omega)$, for (a) granular [$k=870$~N/m, $V=35$~$\mu$m/s, red squares in (a)] and (b) solid friction [$k=$170~N/m, $V=227$~$\mu$m/s, white circles in (c)]. 
%Note that for large $(A\omega)$ in solid friction, the stick-slip amplitude decreases but never vanishes, although in (3) and (4), the slider does not experience any more stick-slip motion but rather a continuous sliding with force fluctuations [the dashed line represents the average of the signal, and the vertical line is the scale for all the data].
}
\end{figure} 

{\it Weakening and suppression of granular or solid friction --}
The pulling velocity $V$, slider mass $m$ and spring stiffness $k$ (Fig.~\ref{fig:expsetup}) are chosen such that, in absence of vibrations, the slider always experiences a well-defined stick-slip motion, characterized by a saw-tooth shape of the instantaneous force applied to the slider (upper signal, insets Figs.~\ref{fig:DF}a,b). At fixed vibration frequency $f=\omega/2\pi$, when increasing the vibration amplitude, we observe a decrease of the normalized force signal amplitude, $\Delta F^*$ (insets Figs.~\ref{fig:DF}a,b). Previous results on granular friction pointed out the vibration velocity, $(A\omega)$, as the governing parameter driving frictional weakening. Reanalysing data from Lastakowski et al.~\cite{Lastakowski15}, we evidence this dependence on the stick-slip amplitude decay (Fig.~\ref{fig:DF}a). Note the convex shape of the curve, independent of the pulling velocity as long as it is small enough for the system to be in stick-slip motion without vibrations. The well-marked transition, at a critical vibration velocity $(A\omega)_c \sim 100$~$\mu$m/s, was found independent of most experimental parameters (grain shape and material, pulling velocity, granular layer thickness, etc.)~\cite{Lastakowski15}.

For solid friction, we also report a weakening and suppression of friction when increasing the vibration amplitude (Fig.~\ref{fig:DF}b). The vibration velocity, $(A\omega)$, is also the governing parameter for the friction decay and suppression, independently of the pulling velocity $V$. Interestingly, some differences appear between solid and granular friction. For Inapa paper, the shape of the frictional weakening is concave, and does not display a clear transition between stick-slip and continuous sliding as in granular assemblies. The gray zone, which indicates the region where stick-slip motion occurs, is delimitated by checking the force signals (Fig.~\ref{fig:DF}b, inset) and picking the vibration velocity above which the slider does not experience any more rest phases, {\it i.e.} linear increases in the force signal. The critical vibration velocity, although of the same order of magnitude than for granular friction, has increased by a factor 2 for Inapa paper, $(A\omega)_c \simeq 200$~$\mu$m/s.

%%%%%%%%%%%%%%%%%%%%%%%%%%%%%%%%%%%%
% FIGURE 3
%%%%%%%%%%%%%%%%%%%%%%%%%%%%%%%%%%%%
\begin{figure}[!t]
\begin{center}
\includegraphics[width=0.8\columnwidth]{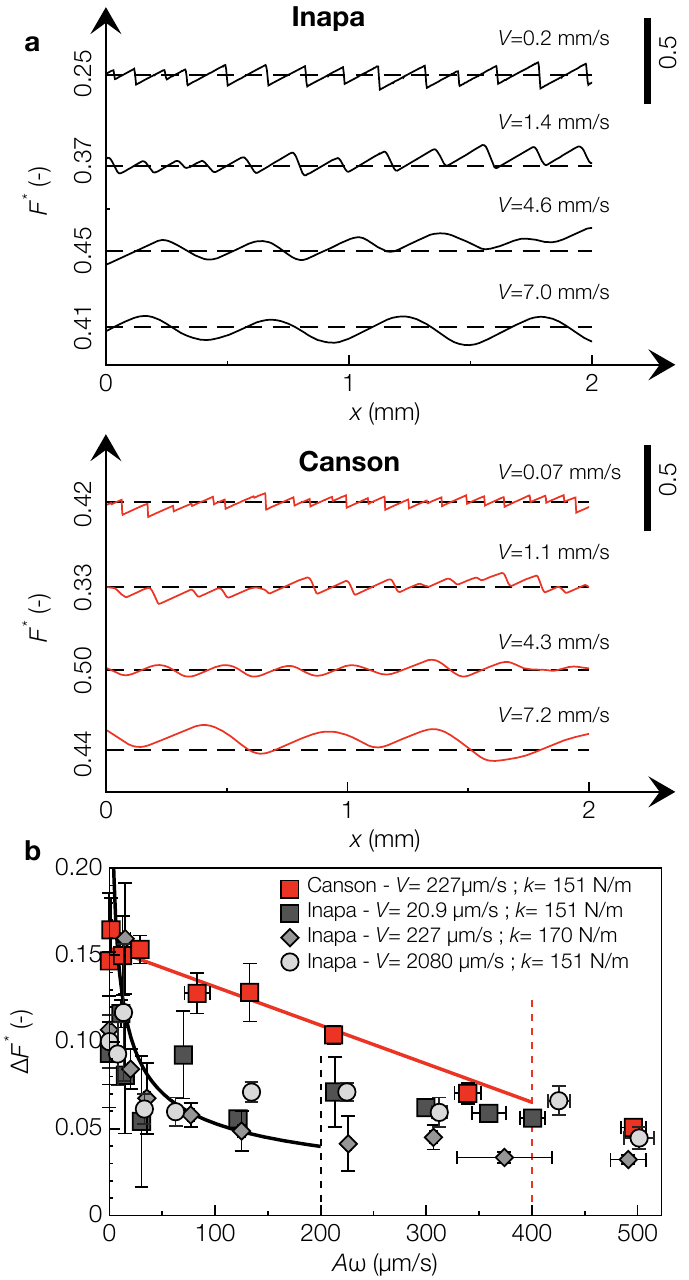}
\end{center}
\caption{\label{fig:VCanson}
Solid (paper-paper) friction dependence on pulling velocity or vibration velocity.
(a) Normalized force $F^*$ as a function of time for different pulling velocity $V$, in absence of vibrations [{\it up,} Inapa paper; {\it down,} Canson$^\circledR$ paper; $k=170$~N/m].
The label on the vertical axes indicate the average value of the dimensionless force [dashed line]. 
(b) Normalized force variations amplitude $\Delta F^*$ for Inapa (gray and black) and Canson$^\circledR$  (red) paper as a function of the vibration velocity, $(A\omega)$
%The Canson$^\circledR$ paper displays a linear decay and a disappearance of stick-slip motion at $(A\omega)_c \simeq 400$~$\mu$m/s (vertical dashed red line), larger than for the printer paper (vertical dashed black line)
[solid lines are guide to the eye, $f=300$~Hz]. 
}
\end{figure}

{\it Role of roughness in frictional weakening --} 
First, the critical vibration velocity is compared to the slider velocity necessary to undergo the transition between stick-slip and continuous sliding in absence of vibrations. Fig.~\ref{fig:VCanson}a (up) displays the normalized force exerted on the slider as a function of time for different pulling velocity $V$, in absence of vibrations $(A\omega)=0$. As predicted by a simple friction model~\cite{Baumberger06}, the slider first undergoes a transition between stick-slip motion and an inertial regime, characterized by a periodic motion without any more rest phases and a relatively large amplitude. However, no steady-sliding is observed in the experimental range of parameters, as it would require a much larger pulling velocity, which cannot be reached by our setup. Therefore, a simple force (or velocity) composition has to be discarded to explain the transition between stick-slip and continuous sliding when imposing vibration to the system.
Second, the critical vibration velocity $(A\omega)_c$ to enter the steady sliding regime in granular assembly was previously explained as the critical energy to overcome a potential energy barrier, namely the typical size of an asperity at the grain surface~\cite{Lastakowski15}. This explanation was in agreement with the independence of $(A\omega)_c$ on the grain shape or material ({\it e.g.} spherical glass beads or irregular ceramic beads), as the typical subscale of roughness was similar, at the nanometer scale.

To test this hypothesis, we performed additional experiments on Canson$^\circledR$, a commercial paper which exhibits a rough surface traditionally used for charcoal drawing. We checked that for this paper, the pulling velocity alone is never enough to provoke the transition between stick-slip motion and continuous sliding (Fig.~\ref{fig:VCanson}a, down), as for Inapa paper. We then investigated the frictional weakening when imposing mechanical disturbances. One again, all data collapse when plotting the force variations amplitude, $\Delta F^*$, as a function of the vibration velocity, $A\omega$ (Fig.~\ref{fig:VCanson}b). However, we observe a change in the critical vibration velocity to undergo the continuous sliding motion. For Canson$^\circledR$ paper, indeed, $(A\omega)_c \simeq 400$~$\mu$m/s (vertical red dashed line, Fig.~\ref{fig:VCanson}b), representing an increase of about a factor 2 respect to Inapa paper. Note that similarly to granular assemblies, both the shape of the frictional weakening curve and the critical vibration velocity are robust for different spring constant $k$ and pulling speed $V$ for a given paper-paper friction (Fig.~\ref{fig:VCanson}b). Changing the paper surface properties, however, induces a change in the curve shape, which exhibits a linear decrease for Canson$^\circledR$.

%%%%%%%%%%%%%%%%%%%%%%%%%%%%%%%%%%%%
% FIGURE 4
%%%%%%%%%%%%%%%%%%%%%%%%%%%%%%%%%%%%
\begin{figure}[!t]
\begin{center}
\includegraphics[width=0.8\columnwidth]{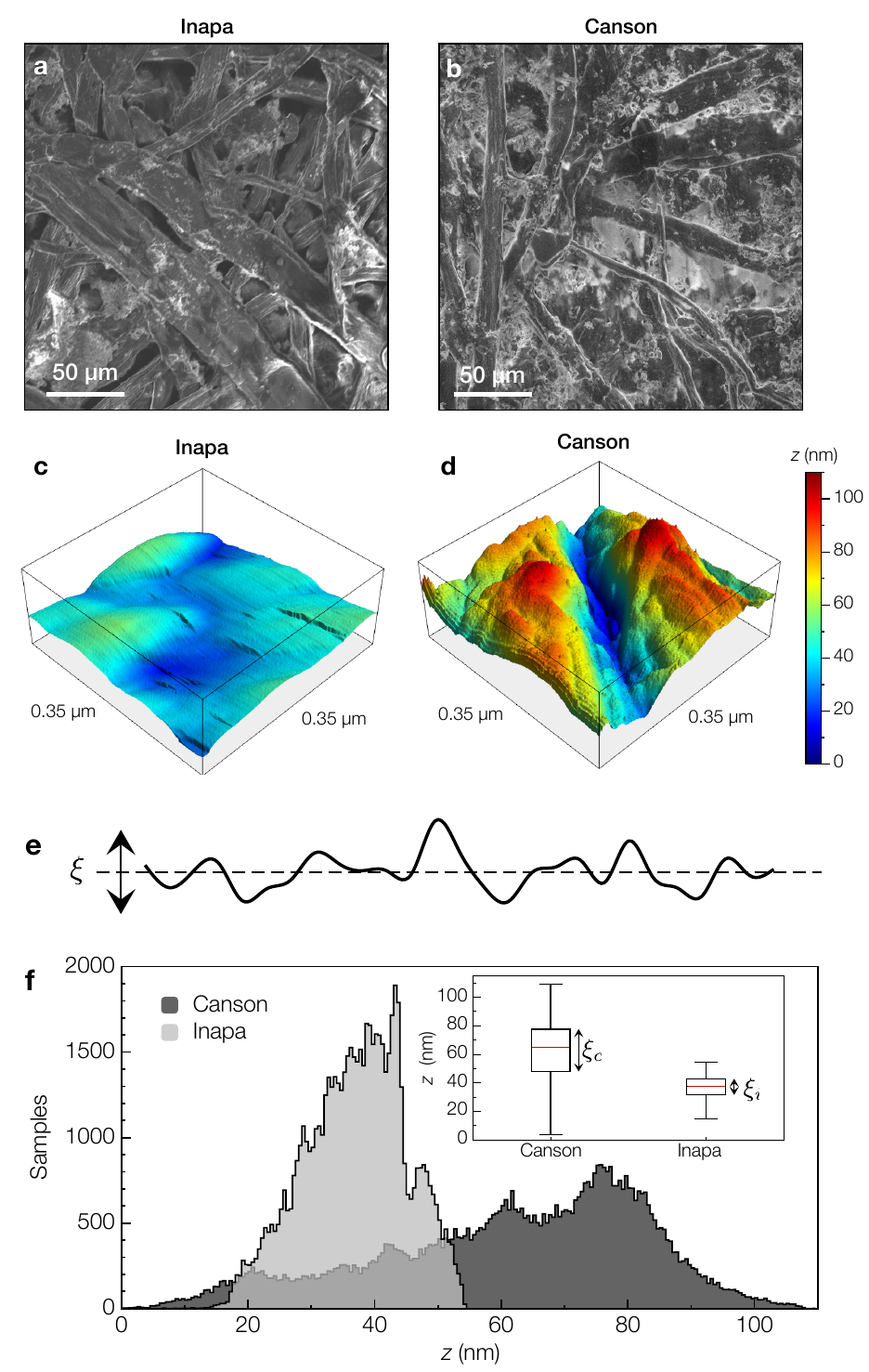}
\end{center}
\caption{\label{fig:AFM} 
Surface roughness analysis.
(a,b) SEM images for Canson$^\circledR$ (a) and Inapa (b) papers. %Note the fibers are larger for Inapa paper, although the paper surface is smoother.
(c,d) AFM topographic map of 0.35~$\mu$m $\times$ 0.35~$\mu$m samples of Canson$^\circledR$ (a) and Inapa (b) papers. The colorbar indicates the height in nanometers.
(e) Schematic view of the dispersion for the smaller scale of roughness. The typical height of asperities is noted $\xi$.
(f) Statistical distribution of the asperities size. {\it Inset:} Box plot representation. The box size represents the IQR (InterQuartile Range) and give the typical size of the asperities for the Canson$^\circledR$, $\xi_c$, and Inapa, $\xi_i$, papers.}
\end{figure}

To further quantify the link with surface roughness properties, we performed Scanning Electron Microscopy (SEM, Supra 55, VP Zeiss) and Atomic Force Microscopy (AFM, NanoWizard$^\circledR$ 4, JPK Instruments) measurements on Inapa and Canson$^\circledR$ paper samples. Figures~\ref{fig:AFM}a,b display SEM images of Inapa (Fig.~\ref{fig:AFM}a) and Canson$^\circledR$ (Fig.~\ref{fig:AFM}b) paper, respectively. Although Canson$^\circledR$ has a rougher surface, paper fibers are smaller than Inapa paper. This indicates that the paper fibers size is not the right scale of roughness, as they are most probably flattened during paper manufacturing. As already proposed for granular assemblies~\cite{Lastakowski15}, the smaller scale of roughness may be the one controlling the friction. Smaller structures can be spotted on Figures~\ref{fig:AFM}a,b, which size is smaller than a few microns. To quantify the smaller scale of roughness, we then performed AFM measurements on both paper samples. Their topography at small scale is clearly different, with rough linear structures for Canson$^\circledR$ (Fig.~\ref{fig:AFM}d) in contrast to smoother bumps for Inapa (Fig.~\ref{fig:AFM}c). The typical scale of the asperities, $\xi$ (Fig.~\ref{fig:AFM}e) is represented in the inset of Figure~\ref{fig:AFM}f for Canson$^\circledR$ ($\xi_c$) and Inapa ($\xi_i$) by the vertical size of the boxes in a box plot representation. We find $\xi_c \simeq 36.8$~nm and $\xi_i \simeq 11.3$~nm.
Following Lastakowski et al.~\cite{Lastakowski15} and DeGiuli \& Wyart~\cite{DeGiuli17}, the energy of the external vibrations necessary to undergo the transition between stick-slip and continuous sliding should be of order $E_e \sim (A\omega)^2$, and comparable to the potential energy barrier to overcome, $E_p \sim \xi$. The critical velocity, $(A\omega)_c$, should therefore scale as the square root of the roughness, $\sqrt{\xi}$. Statistical analysis of the AFM measurements give the ratio between Canson$^\circledR$ and Inapa paper samples, $\sqrt{\xi_c / \xi_i} \sim 1.8$, in agreement with the critical velocity ratio of about 2 for both solid substrates. In the same way, it also gives the typical asperity scale for the grains, $\xi_g = \xi_i / 4 \simeq 2.8$~nm, in agreement with  Lastakowski et al.~\cite{Lastakowski15}, who mentionned the typical size of an asperity at the grains surface to be of the order of a few nanometers.

%{\it Discussion and conclusion --}
{\it Discussion --}
Numerical models for solid friction under vertical vibrations predicted a transition between stick-slip and continuous sliding motion governed by the vibration acceleration, with a threshold at $A\omega^2 \simeq g$, the gravitational acceleration ~\cite{Giacco12}. The above results demonstrate that, as previously shown in granular assemblies, the vibration {\it velocity}, and not acceleration, is the parameter controlling the transition. It is interesting to note that, for the typical frequency range used in the experiment, the threshold velocity corresponds to a very small value of the acceleration, typically a few percents of the gravitational acceleration.

For granular assemblies or, more generally, soft glassy materials, generic models have been proposed to predict the disappearance of the yield stress and trigger of a continuous motion or ``flow'' of the system~\cite{Sollich97}. On the one hand, the classical trap model describes a particle in an energy landscape, where the external mechanical noise acts as an activation mechanism and has an equivalent in terms of effective temperature, even for athermal systems such as granular media~\cite{Bouchaud92,Monthus96,Sollich98,Bocquet09}. On the other hand, more recent models pointed out different statistical behaviours for thermal and athermal systems, and suggested that the mechanical noise would lead to a global inclination of the energy landscape, rather than an increase of the particle energy to escape the well in which it is trapped ~\cite{Nicolas14,Agoritsas15}. Recent experimental evidences on dry granular media have shown that small controlled mechanical fluctuations, whose amplitude is much smaller than the granular assembly yield stress, are enough to provoke a macroscopic flow by an accumulative process: tiny effects integrated over time can lead to the system fluidization, as a secular drift mechanism ~\cite{Derec01,Pons15}. However, although the global effective rheology depends also, in this case, on the product between the vibration amplitude and the frequency - in other words, on the vibration velocity, the authors do not find any velocity threshold $(A\omega)_c$ as their system flows continuously for any tiny applied vibration. A possible explanation could be the existence of a smaller, microscopic scale in the energy landscape basins, as recently suggested by Charbonneau et al.~\cite{Charbonneau14}, which were not captured by the previous experimental devices but successfully captured by our experiments.

Conversely, both numerical ~\cite{Ferdowsi14} and theoretical ~\cite{DeGiuli17} works mention the existence of a vibration threshold below which no frictional weakening occur in granular assemblies, in agreement with laboratory fault-gouge experiments~\cite{Johnson05,Johnson12,Xia13} and field measurements~\cite{Gomberg01}. Both high-pressure experiments and field data suggest an increase of this threshold velocity with the normal load - of the order of 1~cm/s for 0.1~MPa and 10~cm/s for 10~MPa fault load, for instance~\cite{Gomberg01}.

In summary, one could suggest the existence of two thresholds: (1) a ``high'' threshold $(A\omega)_c$ above which the yield stress vanishes and the system experiences a continuous motion; (2) a ``low'''' threshold $(A\omega)^*$ to trigger the flow, increasing with confining pressure - which was not captured by our experimental device. The difficulty in capturing both thresholds lies in measuring tiny effects, at the scale of asperities, together by increasing strongly the normal load (from 0.1 to 10~MPa typically in fault gouges experiments or in the field~\cite{Gomberg01,Johnson05,Johnson08,Xia13}). The experimental challenge is still open.

{\it Acknowledgements --}
The authors thank Ludovic Bellon and Vincent Dolique for their help on SEM and AFM measurements. G.V. acknowledges financial support from PUCV DI Regular No.~039.438/2017. This work was supported by Programa de Cooperaci\'on Cient\'ifica ECOS/CONICYT C14E07 and the Laboratoire International Associ\'e ``Mati\`ere: Structure et Dynamique'' (LIA-MSD, France-Chile).

%%%%%%%%%%%%%%%%%%%%%%%%%%%%%%%%%%%%%%%%%
% BIBLIOGRAPHY
%%%%%%%%%%%%%%%%%%%%%%%%%%%%%%%%%%%%%%%%%

%% Put the bibliography here, most people will use BiBTeX in
%% which case the environment below should be replaced with
%% the \bibliography{} command.

%
%\bibliographystyle{naturemag}
%\bibliography{../../../../BIBLIO/journaux,../../../../BIBLIO/references}
%\bibliography{references}

%\section{Competing Interests}
%The authors declare that they have no competing financial interests.

%\section{Correspondence}
%$^*$Correspondence and requests for materials should be addressed to V.V.~(email: valerie.vidal@ens-lyon.fr).

%%
%% TABLES
%%
%% If there are any tables, put them here.
%%

\end{document}